\documentclass{article}

\usepackage[T1]{fontenc}
\usepackage[utf8]{inputenc}
\usepackage{geometry}
\usepackage{amsmath}
\usepackage{amsthm}
\usepackage{graphicx}
\usepackage{ae,lmodern}
\usepackage{caption}
\usepackage{subcaption}
\usepackage[authoryear]{natbib}
\usepackage{amssymb}
\usepackage{authblk}
\usepackage{tikz}
\usepackage{diagbox}
\usepackage{float}
\usepackage{caption}
\usepackage{verbatim}
\usepackage[colorinlistoftodos]{todonotes}
\usepackage[colorlinks=true, allcolors=blue]{hyperref}

\usepackage{lineno}
\usepackage{setspace}
\usetikzlibrary{matrix,shapes,shapes.arrows,arrows.meta,positioning}

\usetikzlibrary{shapes,shadows,arrows}
\usetikzlibrary{positioning}

\usepackage[normalem]{ulem}

\title{No sensitivity to functional forms in the Rosenzweig-MacArthur model with strong environmental stochasticity}
\newcommand{\cmmnt}[1]{}

\author{Frédéric Barraquand}
\affil{\normalsize Institute of Mathematics of Bordeaux, CNRS \& University of Bordeaux, Talence, France}
\date{}

\begin{document}

\maketitle
\thispagestyle{empty}

\begin{abstract}
\normalsize
The classic Rosenzweig-MacArthur predator-prey model has been shown to exhibit, like other coupled nonlinear ordinary differential equations (ODEs) from ecology, worrying sensitivity to model structure. This sensitivity manifests as markedly different community dynamics arising from saturating functional responses with nearly identical shapes but different mathematical expressions. Using a stochastic differential equation (SDE) version of the Rosenzweig-MacArthur model with the three functional responses considered by Fussmann \& Blasius (2005), I show that such sensitivity seems to be solely a property of ODEs or stochastic systems with weak noise. SDEs with strong environmental noise have by contrast very similar fluctuation patterns, irrespective of the mathematical formula used. Although eigenvalues of linearised predator-prey models have been used as an argument for structural sensitivity, they can also be an argument against structural sensitivity. While the sign of the eigenvalues' real part is sensitive to model structure, its magnitude and the presence of imaginary parts are not, which suggests noise-driven oscillations for a broad range of carrying capacities. I then discuss multiple ways to evaluate structural sensitivity in a stochastic setting, for predator-prey or other ecological systems. 
\end{abstract}

\vspace{1cm}

~\\ 
Correspondence to \url{frederic.barraquand@u-bordeaux.fr}\\
Published in \textit{Journal of Theoretical Biology} with \verb|doi:10.1016/j.jtbi.2023.111566|
\clearpage
\clearpage

\newpage

\section*{Introduction}

Structural sensitivity, as defined by \cmmnt{Wood and Thomas }\citet{wood1999super} and \cmmnt{Fussmann and Blasius }\citet{fussmann2005community}, occurs when small changes in model structure can induce large changes in model behaviour. As \cmmnt{Fussmann and Blasius}\citet{fussmann2005community} demonstrated for the continuous-time predator-prey model of \citet{rosenzweig1963graphical}, picking a particular equation for the functional response (i.e., the kill rate of individual predators as a function of prey density), out of a set of functions that are barely different to the eye, can result in the long run in either a limit cycle or a stable fixed point. Surprisingly, the phenomenon can occur for sizeable sections of parameter space, which makes us reflect upon the arbitrary choices made by modellers: the commonly used Holling (also called Michaelis-Menten) form, for instance, always tends to generate cycles more easily. Structural sensitivity has been judged to be one of the main limits to the predictive power of dynamic models \citep{fussmann2005community,adamson2013can}, and it has even been suggested that such sensitivity may reflect the inherent complexity of ecological processes \citep{adamson2014defining,adamson2016quantifying}. Several mathematical approaches have emerged to both detect such structural sensitivity of nonlinear ordinary differential equations \citep{adamson2013can,adamson2014bifurcation,adamson2014defining}, and mitigate its consequences, by averaging the predictions of multiple plausible models \citep{adamson2016quantifying,aldebert2018structural}. 

The perspective adopted here consists instead of viewing structural sensitivity from a stochastic standpoint. 
\citet{nisbet1976simple,nisbet1982modelling} stressed early on that an underdamped
oscillator sufficiently perturbed by noise could exhibit quasi-cycles, i.e., noise-sustained oscillations with a dominant periodicity. There is therefore reason to believe that an oscillator that presents a stable focus with a pair of complex eigenvalues (instead of a stable node, which only presents real eigenvalues) will often show cycles in a stochastic environment. Much of the research on quasi-cycles has focused on the influence of demographic stochasticity \citep{mckane2004smp,mckane2005ppc,pinedakrch2007ttc}, which means relatively small stochastic perturbations in large populations. With demographic perturbations, effects of noise will certainly be visible, but they may be somewhat limited, and consequently not strong enough to counteract the sensitivity of deterministic models to functional forms---unless populations are very small. This creates an incentive to investigate what happens in the presence of large environmental perturbations, that are also believed to dominate the stochasticity experienced by most large natural populations \citep{lande2003stochastic,mutshinda2009drives}. An overlooked phenomenon in random dynamical systems is that of ``noisy precursors'' \citep{Wiesenfeld1985Noisy}: close to and below a supercritical Hopf bifurcation in the deterministic skeleton of the model, the behaviour of its stochastic counterpart will make it look like the bifurcation has already been crossed \citep{Wiesenfeld1985Noisy,neiman1997coherence,wyse2022structural}. Or, in other words, environmental noise smoothes over the transition between fixed point stability and cyclicity. This also suggests that strong stochasticity may remove some of the near-pathological behaviour of deterministic models. 

In the following, I introduce strong environmental stochasticity in the population growth rates of prey and predator in a stochastic differential equation (SDE) version of the Rosenzweig-MacArthur model \citep{rosenzweig1963graphical,rosenzweig1971ped}. I show numerically that the amplitude of cycles reacts to the bifurcation parameter $K$ in smooth and similar manner for the three functional responses considered by \citet{fussmann2005community}, so that the sensitivity to structure is essentially removed from the (strongly) stochastic models. A reinterpretation of how real and imaginary eigenvalues of the deterministic skeleton of the model change with $K$ further illustrates why such stochasticity smoothes dynamical transitions and largely forbids structural sensitivity.  

\section*{Model and methods}

The stochastic Rosenzweig-MacArthur (RMA) model is formulated as a system of coupled stochastic differential equations (SDEs) in It\^o form, defined by eqs.~\ref{eq:SDE_RMA_prey}--\ref{eq:SDE_RMA_pred} for the prey ($X$) and the predator ($Y$) densities, that are random variables evolving over time:

\begin{align}
    dX = {}& \left( g(X) - f(X) Y \right) dt + \sigma_E X dW_x\label{eq:SDE_RMA_prey}\\
    dY = {}& \left( \epsilon f(X) Y - m Y \right)dt + \sigma_E Y dW_y. 
    \label{eq:SDE_RMA_pred}
\end{align}

We use in this article the exact same deterministic skeleton as \cmmnt{Fussmann and Blasius }\citet{fussmann2005community}, with three functional forms for the functional response, $f_H(x) = \frac{a_H x}{1+b_H x}$,$f_I(x) = a_I\left(1-\exp(-b_I x)\right)$, $f_T(x) = a_T \tanh(b_T x)$ and a logistic prey growth rate function $g(x)=r x\left(1-\frac{x}{K}\right)$. The functional response parameters are identical to \cmmnt{Fussmann and Blasius }\citet{fussmann2005community}, $a_H = 3.05,b_H = 2.68$; $a_I = 1,b_I = 2$; $a_T = 0.99,b_T = 1.48$. We also use the same other parameters values $r=1,\epsilon = 1, m=0.1$ (the conversion efficiency was added to be able to also consider more realistic $\epsilon<1$ values, though we stick to $\epsilon=1$ for main text simulations). $W_x(t)$ and $W_y(t)$ are uncorrelated Wiener processes. The environmental noise standard deviation is set to $\sigma_E=0.25$ ($\sigma_E=0.1$ is considered in \ref{app:add_bif_diags}). 

The numerical integration is performed using the simplest possible Euler-Maruyama scheme, with steps of size $\delta t = 0.05$ for $K \in [0, 1.5]$, with integration over $2000$ timesteps, and $\delta t = 0.005$ for $K \in [0, 15]$ (\ref{app:add_bif_diags}). The ODE integration uses an Euler scheme for the same step size; Matlab's \verb|ode45| provides a good match, suggesting that it is a sufficiently small. The eigenvalues of the nontrivial fixed point of the deterministic model have been computed through an analytical expression of the Jacobian, computed numerically at the equilibrium for the chosen parameter values. The computer code is available at \url{https://github.com/fbarraquand/sensitivity_stochRMA}. The original code, used for the figures presented here, has been written and executed in Matlab (R2017a). As always with numerical results, there is a risk of a different kind of sensitivity to the computational implementation of the models. Therefore, the main result of Fig~\ref{fig:RMA_Kloop} has been replicated in C++ for some parameter values (see code repository). 

\section*{Results}

\begin{figure}[ht]
    \centering
    \includegraphics[width=\textwidth]{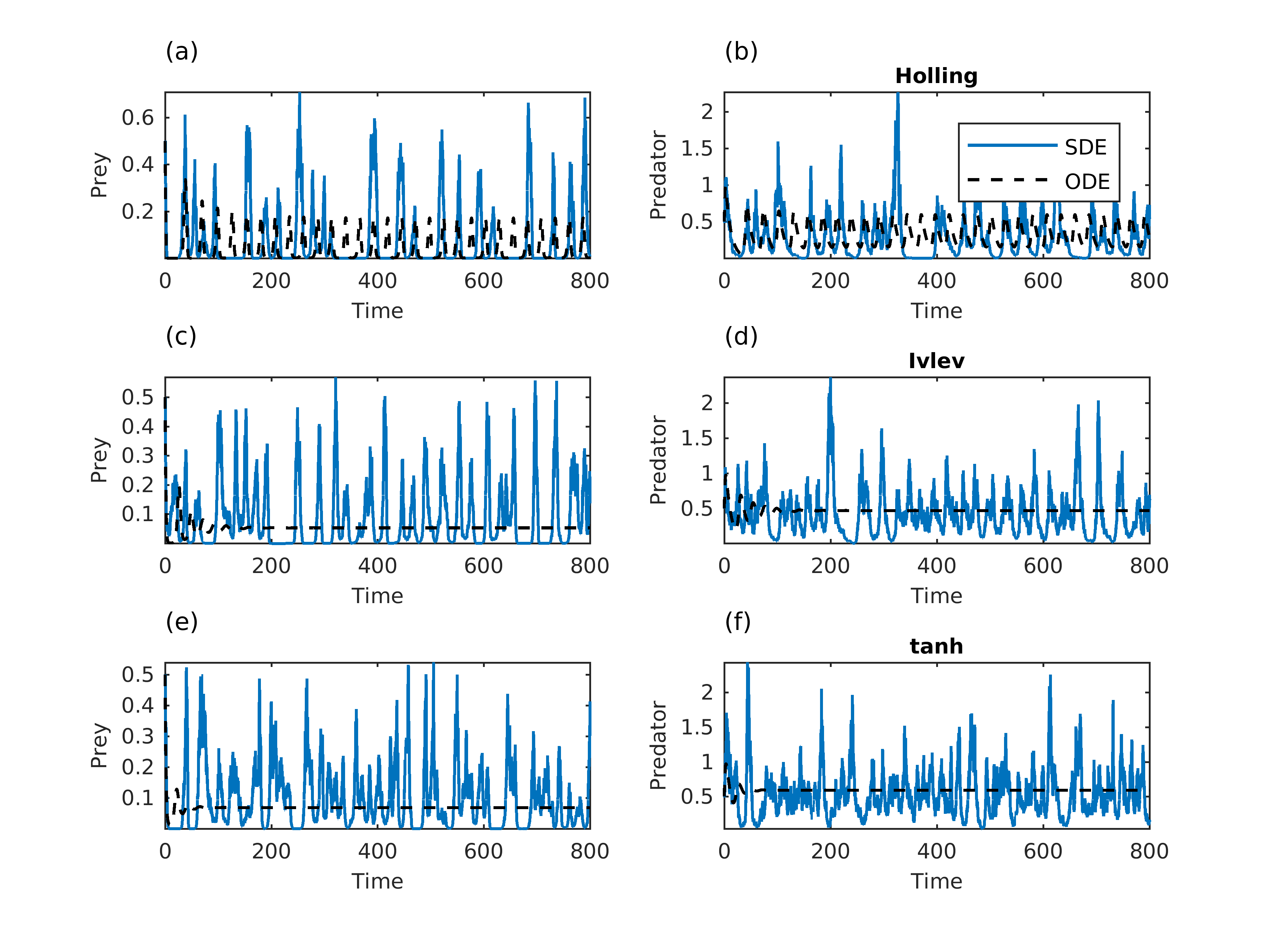}
    \caption{Trajectories of the ODEs (dashed black lines) and SDEs (plain blue lines) for the 3 functional responses. Top (a-b), Holling; middle (c-d), Ivlev; bottom (e-f), tanh. Functional response parameters are given in the text. Other parameters $r = 1, \; K = 0.5, \; \epsilon = 1, \; m = 0.1 \;, \sigma_E = 0.25$. \label{fig:RMA_Traj_K0.5}}
\end{figure}

Comparisons of the deterministic and stochastic trajectories (example for $K=0.5$ in Fig.~\ref{fig:RMA_Traj_K0.5}) reveal that while the dynamical behaviour of the ODEs for Holling, Ivlev and tanh functional responses can be markedly different, the stochastic trajectories are not too dissimilar. For this value of $K$, only the Holling functional response model would have crossed the Hopf, with relatively mild-amplitude oscillations (Fig.~\ref{fig:RMA_Traj_K0.5}a,b). All three models show wide-amplitude oscillations though. This can be related to oscillatory transients in the ODEs, that are excited by noise. In other words, oscillatory trajectories that would converge to a stable focus in a deterministic setup (dashed lines in Fig.~\ref{fig:RMA_Traj_K0.5}c,d,e,f) are recurrently perturbed under environmental stochasticity, which favours sustained, broad-amplitude oscillations. 

\begin{figure}[ht]
    \centering
    \includegraphics[width=\textwidth]{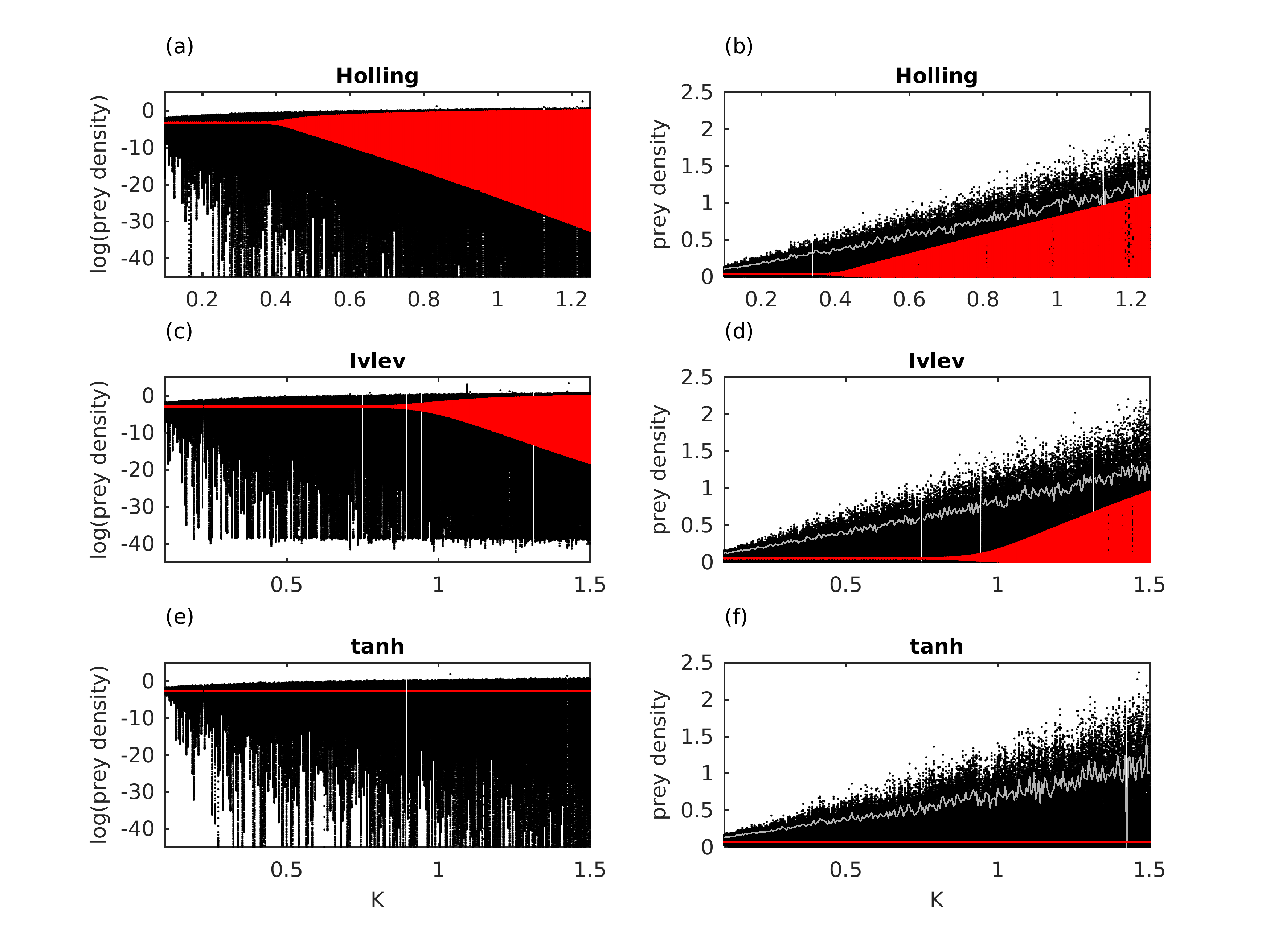}
    \caption{Bifurcation diagrams for the ODEs (red dots) and SDEs (black dots). The grey line is the 97.5th percentile. Top (a-b), Holling; middle (c-d), Ivlev; bottom (e-f), tanh functional responses. Left column, logarithmic scale; right column, untransformed scale. Functional response parameters are given in the text. Other parameters $r = 1, \; \epsilon = 1, \; m = 0.1 \;, \sigma_E = 0.25$.}
    \label{fig:RMA_Kloop}
\end{figure}

The deterministic and stochastic versions of the bifurcation diagram with respect to the carrying capacity $K$ reveal that while structural sensitivity is very present in the deterministic model, it disappears almost entirely in the stochastic versions (Fig.~\ref{fig:RMA_Kloop}). SDEs with a little less environmental noise still show some differences in the stochastic bifurcation diagrams (\ref{app:add_bif_diags}), but these differences are much fainter than in the deterministic model. These results are also robust to lower values of $\epsilon$ (see code repository), and the power spectra are similar for the three functional responses under strong noise (\ref{app:power_spectrum}). 

\begin{figure}[ht]
    \centering
    \includegraphics[width=0.65\textwidth]{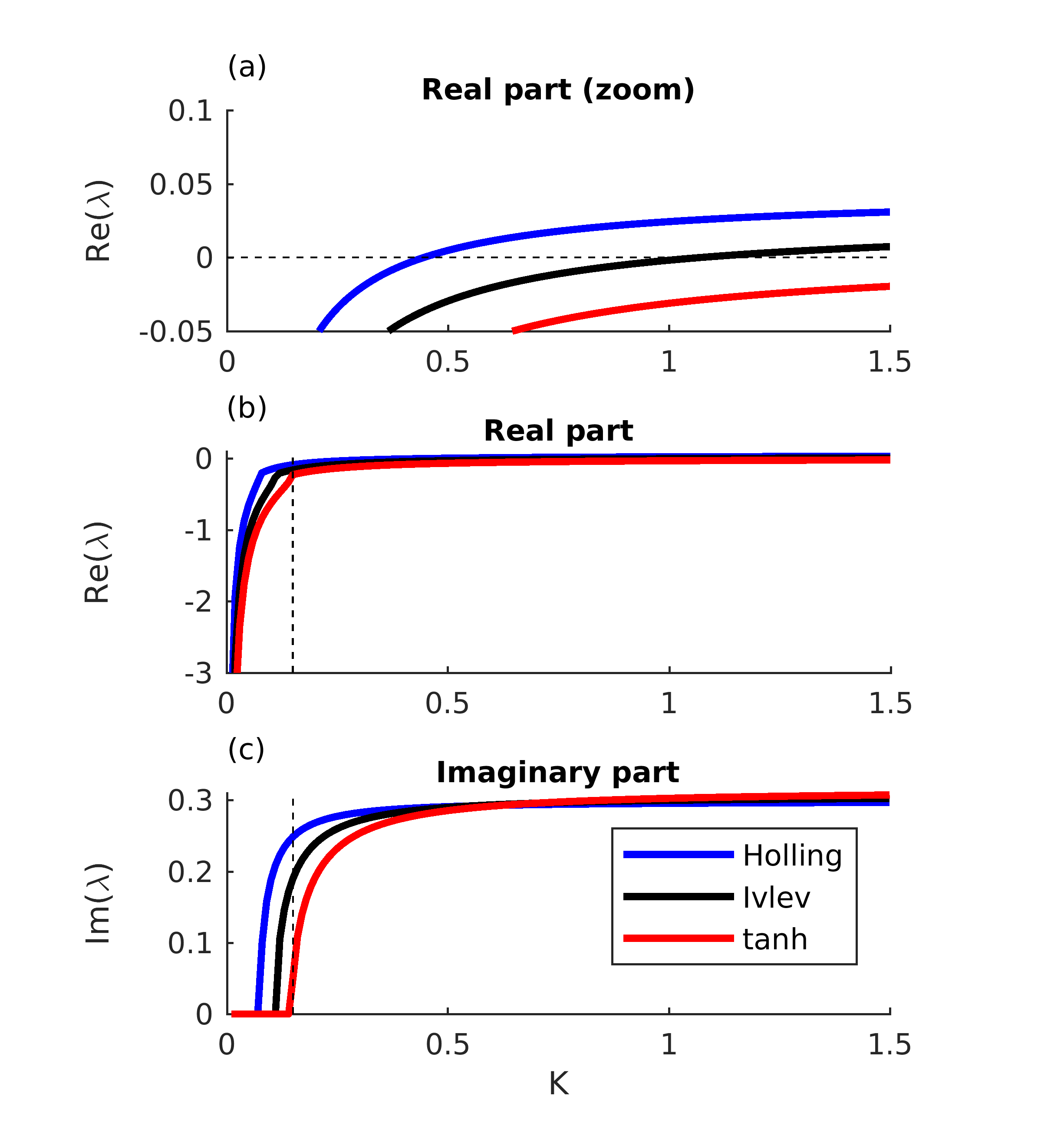}
    \caption{Eigenvalues of the nontrivial fixed point for the three functional responses, as a function of the carrying capacity $K$. Panel (a) shows for which functional responses the real part of the dominant eigenvalue is positive, as  presented by Fussmann and Blasius (2005); panel (b) zooms out to show the lowest eigenvalues for very low $K$, the vertical dashed line separating fast from slow return times to equilibrium; (c) presents the imaginary part of the eigenvalue. In (c), the vertical dashed line materializes the $K$ value for which eigenvalues associated to all functional responses are complex.}
    \label{fig:eigenvalues}
\end{figure}

Despite the apparent increased complexity of stochastic models, these differences between deterministic and stochastic models can be readily explained by the eigenvalues of the linearized model at the positive fixed point. In the case of a focus (oscillatory convergence or divergence to the fixed point), the solutions of the linearized ODEs and SDEs are proportional to $\exp(\mu t \pm i\omega t)$ where $\mu$ is the real part and $\omega$ is the imaginary part of the eigenvalues. When presenting how deterministic structural sensitivity can arise, \cmmnt{Fussmann and Blasius }\citet{fussmann2005community}, in their Fig. 2, zoomed in on the section of $\text{Re}(\lambda)$ values ($\mu$ here, $\tau$ in \citealt{fussmann2005community}) that crosses zero. In a deterministic and asymptotic perspective this makes sense: one wants to pinpoint exactly at which $K$ value the fixed point becomes a limit cycle and vice versa. They noticed that the zero crossing for $\text{Re}(\lambda)$ was obtained at quite different $K$ values for the various functional forms (reproduced in Fig~\ref{fig:eigenvalues}a here, though \citet{fussmann2005community} consider also larger $K$ values). However, in a stochastic setting we need to think not only of the sign of $\text{Re}(\lambda)$ but also its magnitude. If we change the y-axis scale of $\text{Re}(\lambda)$ and, unlike \citet{fussmann2005community}, decide to consider all observed $\text{Re}(\lambda)$ values from $K=0$ to $1.5$ (Fig ~\ref{fig:eigenvalues}b), we see that (i) $\text{Re}(\lambda)$, even if negative, stops to be a main pulling force to the equilibrium for relatively low $K$ ($\approx 0.2$, Fig ~\ref{fig:eigenvalues}b), long before the bifurcation points. Indeed, the return time after perturbation being of the order $\frac{1}{\text{Re}(\lambda)}$, we can deduce much longer return times for $K>0.2$.  Considering as well the imaginary part (Fig ~\ref{fig:eigenvalues}c), $\text{Im}(\lambda)$ emerges as nonzero for very low $K$ as well ($K<0.2$)---and nonzero $\text{Im}(\lambda)$ means oscillations. Combining information on both $\text{Re}(\lambda)$ and $\text{Im}(\lambda)$, we conclude that there is a broad range of $K$ values below the bifurcation point, for all three functional responses, where large noise-sustained oscillations are possible because $\text{Im}(\lambda)$ is nonzero and $\text{Re}(\lambda)$ is very low. 

To sum up, the asymptotic limit cycle behaviour, controlled by $\text{sign}(\text{Re}(\lambda))$, is very sensitive to the functional form (Fig ~\ref{fig:eigenvalues}a). However, the magnitude of the real part (Fig ~\ref{fig:eigenvalues}b), that governs the response time to perturbations, and the imaginary part (Fig ~\ref{fig:eigenvalues}c), that governs the oscillatory tendency, are not that sensitive to $K$. This is what explains why all three stochastic models start to oscillate in a cyclic manner for very low $K$ values, with a smooth increase in amplitude as $K$ increases. 

Let us note one small caveat: the $\tanh$ functional response predator-prey model is a bit more difficult to interpret through eigenvalues of the fixed point, because this assumes a single fixed point and there is a actually multistability \citep{fussmann2005community,seo2018sensitivity} with a coexisting limit cycle for some $K$ values. 

In passing, from Fig.~\ref{fig:eigenvalues} it is easy to deduce the average period of the quasi-cycles, $\text{frequency} = \frac{\omega}{2\pi} \approx \frac{0.3}{2\pi}$, so $\text{period} = 1/\text{frequency} = 20$ time units for most $K$ values here. Overall, the imaginary part of the eigenvalue at the fixed point is as rich in information as the real part. The actual dominant frequencies in the signal are slightly below $1/20$ with strong noise (\ref{app:power_spectrum}).

\section*{Discussion}

Although the asymptotic behaviour of the deterministic Rosenzweig-MacArthur predator-prey model is very sensitive to the type of functional forms used for the functional response, we have seen that such predictions are radically altered when strong environmental stochasticity is introduced in all three versions of this model. The stochastic bifurcations diagrams for similar Holling, Ivlev, and tanh functional responses were indeed difficult to differentiate, in stark contrast to the deterministic ones. 

While the results may appear at first sight surprising because of the large difference between stochastic and deterministic model behaviour, such strong interaction between transients and stochasticity is by no means unexpected. In fact, the idea has been around for decades (see \cmmnt{Piélou 1977 }\citealt{pielou1977mathematical}, p. 109), together with the mathematical machinery to analyse it (see \cmmnt{Nisbet and Gurnet 1982 }\citealt{nisbet1982modelling}, p. 98). 
Yet the workhorse of theoretical ecology seems to remain nonlinear ODEs---given this, it is important to get an interpretation of deterministic stability analyses that is compatible with a stochastic worldview. Two interpretation pitfalls when studying with ODEs (and their linearizations) ecological systems that are actually stochastic are (i) focusing solely on the sign of the real part of the dominant eigenvalue, rather than its actual value (e.g., is $\text{Re}(\lambda_{\text{dom}})$ \emph{very} negative or just below zero?) and (ii) not reporting the imaginary part of the dominant eigenvalues. Indeed, as soon as the imaginary part is nonzero the solution oscillates, and if $\text{Re}(\lambda_{\text{dom}})$ is small the return times will be long, so that the system may oscillate for a very long time and never reach in practice a stable equilibrium. A similar conclusion that a too narrow focus on asymptotic stability of deterministic systems can be misleading can be found on the abundant literature on ecological transients (see e.g. \cmmnt{Hastings et al. }\citealt{hastings2004transients,hastings2018transient}). 

Several authors have noted that stochastic nonlinear dynamical systems tended to ``cross the Hopf bifurcation earlier'' \citep{Wiesenfeld1985Noisy,neiman1997coherence,gellner2016duality,barraquand2017moving,wyse2022structural}, in the sense that the behaviour right after the bifurcation (limit cycles) is emulated by the forced system right before the bifurcation. This was the case here as well.  These ``noisy precursors'' are not exactly the same as the quasi-cycles that would occur in a damped Lotka-Volterra system (where no limit cycle can be reached), but instead constitute an early transition towards noisy limit cycle behaviour, induced by fluctuations in a bifurcation parameter. In the case of a (supercritical) Hopf, this phenomenon is not clear-cut because the transition between the excited oscillatory regime and forced limit cycle regime is relatively smooth (Fig.~\ref{fig:RMA_Kloop}, see also \citealt{dutta2018robustness}). In fact, in models with a supercritical Hopf, the transients become more and more pronounced in the deterministic model as we get near the bifurcation, so that in a stochastic environment, it becomes difficult to differentiate between noise-sustained oscillations from limit cycles without a detailed examination of the power spectrum \citep{louca2014distinguishing}. Other kinds of bifurcations might make noise-induced transitions more evident, such as period-doubling (unless noise destroys the periodicity). Overall, it useful to remember that noise can have effects both relatively far away from bifurcations, by exciting the transients to a deterministic fixed point, and effects close to a bifurcation point, by making the system cross the Hopf bifurcation for different parameter values than in the deterministic setting.  

These results, together with the existing body of stochastic population and community dynamics literature \citep{nisbet1982modelling,nolting2016balls,arnoldi2019inherent}, some of which pertains to finding signals of regime transitions \citep{kefi2013early,dutta2018robustness}, can help us to make suggestions for further work on the structural sensitivity of ecological models. First, environmental noise effects on population and community dynamics are usually larger than anticipated, especially in the case of cycling populations, as recognized early on by \cmmnt{Nisbet and Gurney }\citet{nisbet1976simple}. In case of sensitivity to structure detected in a deterministic model, a healthy check is therefore to introduce (multiplicative) noise in various amounts in the model to verify whether such structural sensitivity is solely a property of the deterministic model or whether it applies to the stochastic model as well \citep{wood1999super}. In the simple example with one predator and one prey that we have considered, structural sensitivity was shown to disappear under strong environmental noise on the population growth rates. This remains partly true when introducing noise elsewhere in the model (see \ref{app:large_noise_FR}), although noise on the functional response's half-saturation constant provides slightly less similar bifurcation diagrams than multiplicative noise on the growth rates. This may explain the difference between this study's main results and \cmmnt{Wyse et al. }\citet{wyse2022structural}, who independently found reduced---but not absent---structural sensitivity in predator-prey models where only the functional response was stochastic. There are, however, a broad range of deterministic models that display asymptotic structural sensitivity \citep{adamson2014defining}. It might well be that some still exhibit structural sensitivity under strong environmental noise.

Second, although qualitative conclusions with ODE models are completely fine (e.g., when increasing $K$ or the attack rate, a Hopf bifurcation eventually occurs), a too heavy focus on the bifurcation structure associated with the \emph{asymptotic deterministic behaviour of the model} might distract modellers from investigating transient behaviours \citep{hastings2004transients,hastings2018transient}. Studying transients that follow a given perturbation, even in a purely deterministic model, is a first step towards taking into account the environmental variation that characterizes real ecosystems \citep[see e.g.][]{arnoldi2018ecosystems}. Investigating more ecologically meaningful structural sensitivity in a ODE framework would probably imply developing (or adapting) metrics to compare transients. Most structural sensitivity approaches focus instead on asymptotic behaviour, though the possibility of studying transients has been mentioned \citep{adamson2014defining}. However, the interactions between transients and noise suggest that SDEs or other random dynamical systems will be paramount to compare the dynamics generated by alternative model specifications in a stochastic world. 

Properly quantifying the difference in model behaviours generated by SDEs with different functional forms is challenging though. An idea to quantify more formally the distance between stochastic attractors could be to use quasi-potential techniques, introduced in ecology by \cmmnt{Nolting and Abbott }\citet{nolting2016balls}, and derived from Freidlin-Wentzell theory. Stochastic differential equations with the environmental (and thus multiplicative) noise considered above may need to be converted first to an additive scale through Itô's lemma \citep{moore2015qpot}, but this poses no problem. \cmmnt{Nolting and Abbott }\citet{nolting2016balls} note, however, that the approximation used to produce the quasi-potential works well for small noise, which differs from the choices made above. A simulation approach might therefore often be part of the answer to study systems affected by strong noise. In the (stochastic) bifurcation diagrams showed in this paper, I have chosen to plot all values of the simulated stochastic process, so that Fig.~\ref{fig:RMA_Kloop} shows the upper and lower bounds for each $K$ value over a long period of time, all the values in-between, and a 97.5th percentile (in grey). The advantage of defining a stochastic bifurcation diagram as such, outside of sheer simplicity, is that the definition would apply equally well to a chaotic system with a strange attractor (such as the logistic or Ricker map in discrete time). Yet, a difference between the deterministic and the stochastic bifurcation diagram is that as time goes to infinity, the minima and maxima of the stochastic process still grow increasingly apart---by contrast deterministic attractors have sharper boundaries. In other words, the tails of the steady-state distribution of the stochastic process modelled by coupled SDEs are increasingly revealed as one considers longer time frames. 
For this reason, it may be that evaluating in detail the structural sensitivity of coupled nonlinear SDEs requires to compare the full steady-state distributions under different model specifications or the moments of such distributions. 

\subsection*{Acknowledgements}
I thank the anonymous referee of a previous publication who kindly pointed towards the literature on noisy precursors of instabilities, and Matthew Adamson for the discussion in 2018 that motivated me to pursue this work, as well as for multiple suggestions to improve it. I also thank Hannes Erbis for an independent replication of some results in 2020, and Frank Hilker and two referees for comments on the manuscript. 

\bibliography{sensitivityRMA}

\begin{thebibliography}{33}
\expandafter\ifx\csname natexlab\endcsname\relax\def\natexlab#1{#1}\fi

\bibitem[{Adamson \& Morozov(2013)}]{adamson2013can}
Adamson, M. \& Morozov, A.Y. (2013).
\newblock When can we trust our model predictions? unearthing structural
  sensitivity in biological systems.
\newblock \emph{Proceedings of the Royal Society A: Mathematical, Physical and
  Engineering Sciences}, 469, 20120500.

\bibitem[{Adamson \& Morozov(2014{\natexlab{a}})}]{adamson2014bifurcation}
Adamson, M. \& Morozov, A.Y. (2014{\natexlab{a}}).
\newblock Bifurcation analysis of models with uncertain function specification:
  how should we proceed?
\newblock \emph{Bulletin of Mathematical Biology}, 76, 1218--1240.

\bibitem[{Adamson \& Morozov(2014{\natexlab{b}})}]{adamson2014defining}
Adamson, M. \& Morozov, A.Y. (2014{\natexlab{b}}).
\newblock Defining and detecting structural sensitivity in biological models:
  developing a new framework.
\newblock \emph{Journal of Mathematical Biology}, 69, 1815--1848.

\bibitem[{Adamson \emph{et~al.}(2016)Adamson, Morozov \&
  Kuzenkov}]{adamson2016quantifying}
Adamson, M., Morozov, A.Y. \& Kuzenkov, O. (2016).
\newblock Quantifying uncertainty in partially specified biological models: how
  can optimal control theory help us?
\newblock \emph{Proceedings of the Royal Society A: Mathematical, Physical and
  Engineering Sciences}, 472, 20150627.

\bibitem[{Aldebert \emph{et~al.}(2018)Aldebert, Kooi, Nerini \&
  Poggiale}]{aldebert2018structural}
Aldebert, C., Kooi, B.W., Nerini, D. \& Poggiale, J.C. (2018).
\newblock {Is structural sensitivity a problem of oversimplified biological
  models? Insights from nested Dynamic Energy Budget models}.
\newblock \emph{Journal of Theoretical Biology}, 448, 1--8.

\bibitem[{Arnoldi \emph{et~al.}(2018)Arnoldi, Bideault, Loreau \&
  Haegeman}]{arnoldi2018ecosystems}
Arnoldi, J.F., Bideault, A., Loreau, M. \& Haegeman, B. (2018).
\newblock How ecosystems recover from pulse perturbations: A theory of short-to
  long-term responses.
\newblock \emph{Journal of Theoretical Biology}, 436, 79--92.

\bibitem[{Arnoldi \emph{et~al.}(2019)Arnoldi, Loreau \&
  Haegeman}]{arnoldi2019inherent}
Arnoldi, J.F., Loreau, M. \& Haegeman, B. (2019).
\newblock The inherent multidimensionality of temporal variability: how common
  and rare species shape stability patterns.
\newblock \emph{Ecology Letters}, 22, 1557--1567.

\bibitem[{Barraquand \& Gimenez(2021)}]{barraquand2021fitting}
Barraquand, F. \& Gimenez, O. (2021).
\newblock Fitting stochastic predator--prey models using both population
  density and kill rate data.
\newblock \emph{Theoretical Population Biology}, 138, 1--27.

\bibitem[{Barraquand \emph{et~al.}(2017)Barraquand, Louca, Abbott, Cobbold,
  Cordoleani, DeAngelis, Elderd, Fox, Greenwood, Hilker
  \emph{et~al.}}]{barraquand2017moving}
Barraquand, F., Louca, S., Abbott, K.C., Cobbold, C.A., Cordoleani, F.,
  DeAngelis, D.L., Elderd, B.D., Fox, J.W., Greenwood, P., Hilker, F.M.
  \emph{et~al.} (2017).
\newblock Moving forward in circles: challenges and opportunities in modelling
  population cycles.
\newblock \emph{Ecology Letters}, 20, 1074--1092.

\bibitem[{Dutta \emph{et~al.}(2018)Dutta, Sharma \&
  Abbott}]{dutta2018robustness}
Dutta, P.S., Sharma, Y. \& Abbott, K.C. (2018).
\newblock Robustness of early warning signals for catastrophic and
  non-catastrophic transitions.
\newblock \emph{Oikos}, 127, 1251--1263.

\bibitem[{Fussmann \& Blasius(2005)}]{fussmann2005community}
Fussmann, G.F. \& Blasius, B. (2005).
\newblock Community response to enrichment is highly sensitive to model
  structure.
\newblock \emph{Biology Letters}, 1, 9--12.

\bibitem[{Gellner \emph{et~al.}(2016)Gellner, McCann \&
  Hastings}]{gellner2016duality}
Gellner, G., McCann, K.S. \& Hastings, A. (2016).
\newblock The duality of stability: towards a stochastic theory of species
  interactions.
\newblock \emph{Theoretical Ecology}, 9, 477--485.

\bibitem[{Hastings(2004)}]{hastings2004transients}
Hastings, A. (2004).
\newblock Transients: the key to long-term ecological understanding?
\newblock \emph{Trends in Ecology \& Evolution}, 19, 39--45.

\bibitem[{Hastings \emph{et~al.}(2018)Hastings, Abbott, Cuddington, Francis,
  Gellner, Lai, Morozov, Petrovskii, Scranton \&
  Zeeman}]{hastings2018transient}
Hastings, A., Abbott, K.C., Cuddington, K., Francis, T., Gellner, G., Lai,
  Y.C., Morozov, A., Petrovskii, S., Scranton, K. \& Zeeman, M.L. (2018).
\newblock Transient phenomena in ecology.
\newblock \emph{Science}, 361, eaat6412.

\bibitem[{K{\'e}fi \emph{et~al.}(2013)K{\'e}fi, Dakos, Scheffer, Van~Nes \&
  Rietkerk}]{kefi2013early}
K{\'e}fi, S., Dakos, V., Scheffer, M., Van~Nes, E.H. \& Rietkerk, M. (2013).
\newblock Early warning signals also precede non-catastrophic transitions.
\newblock \emph{Oikos}, 122, 641--648.

\bibitem[{Lande \emph{et~al.}(2003)Lande, Engen \&
  Saether}]{lande2003stochastic}
Lande, R., Engen, S. \& Saether, B.E. (2003).
\newblock \emph{Stochastic population dynamics in ecology and conservation}.
\newblock Oxford University Press.

\bibitem[{Louca \& Doebeli(2014)}]{louca2014distinguishing}
Louca, S. \& Doebeli, M. (2014).
\newblock Distinguishing intrinsic limit cycles from forced oscillations in
  ecological time series.
\newblock \emph{Theoretical Ecology}, 7, 381--390.

\bibitem[{McKane \& Newman(2004)}]{mckane2004smp}
McKane, A. \& Newman, T. (2004).
\newblock {Stochastic models in population biology and their deterministic
  analogs}.
\newblock \emph{Physical Review E}, 70, 41902.

\bibitem[{McKane \& Newman(2005)}]{mckane2005ppc}
McKane, A. \& Newman, T. (2005).
\newblock {Predator-prey cycles from resonant amplification of demographic
  stochasticity}.
\newblock \emph{Physical Review Letters}, 94, 218102.

\bibitem[{Moore \emph{et~al.}(2015)Moore, Stieha, Nolting, Cameron \&
  Abbott}]{moore2015qpot}
Moore, C.M., Stieha, C.R., Nolting, B.C., Cameron, M.K. \& Abbott, K.C. (2015).
\newblock Qpot: an r package for stochastic differential equation
  quasi-potential analysis.
\newblock \emph{arXiv preprint arXiv:1510.07992}.

\bibitem[{Mutshinda \emph{et~al.}(2009)Mutshinda, O'Hara \&
  Woiwod}]{mutshinda2009drives}
Mutshinda, C.M., O'Hara, R.B. \& Woiwod, I.P. (2009).
\newblock What drives community dynamics?
\newblock \emph{Proceedings of the Royal Society B: Biological Sciences}, 276,
  2923--2929.

\bibitem[{Neiman \emph{et~al.}(1997)Neiman, Saparin \&
  Stone}]{neiman1997coherence}
Neiman, A., Saparin, P.I. \& Stone, L. (1997).
\newblock Coherence resonance at noisy precursors of bifurcations in nonlinear
  dynamical systems.
\newblock \emph{Physical Review E}, 56, 270.

\bibitem[{Nisbet \& Gurney(1976)}]{nisbet1976simple}
Nisbet, R. \& Gurney, W. (1976).
\newblock A simple mechanism for population cycles.
\newblock \emph{Nature}, 263, 319--320.

\bibitem[{Nisbet \& Gurney(1982)}]{nisbet1982modelling}
Nisbet, R. \& Gurney, W. (1982).
\newblock \emph{{Modelling fluctuating populations}}.
\newblock John Wiley \& Sons.

\bibitem[{Nolting \& Abbott(2016)}]{nolting2016balls}
Nolting, B.C. \& Abbott, K.C. (2016).
\newblock Balls, cups, and quasi-potentials: quantifying stability in
  stochastic systems.
\newblock \emph{Ecology}, 97, 850--864.

\bibitem[{Pielou(1977)}]{pielou1977mathematical}
Pielou, E. (1977).
\newblock \emph{Mathematical ecology.[2d ed.].}
\newblock 2nd edn.
\newblock John Wiley and Sons, New York, USA.

\bibitem[{Pineda-Krch \emph{et~al.}(2007)Pineda-Krch, Blok, Dieckmann \&
  Doebeli}]{pinedakrch2007ttc}
Pineda-Krch, M., Blok, J., Dieckmann, U. \& Doebeli, M. (2007).
\newblock {A tale of two cycles-distinguishing quasi-cycles and limit cycles in
  finite predator-prey populations.}
\newblock \emph{Oikos}, 116, 53.

\bibitem[{Rosenzweig(1971)}]{rosenzweig1971ped}
Rosenzweig, M. (1971).
\newblock {Paradox of enrichment: destabilization of exploitation ecosystems in
  ecological time}.
\newblock \emph{Science}, 171, 385--387.

\bibitem[{Rosenzweig \& MacArthur(1963)}]{rosenzweig1963graphical}
Rosenzweig, M. \& MacArthur, R. (1963).
\newblock {Graphical representation and stability conditions of predator-prey
  interactions}.
\newblock \emph{American Naturalist}, 97, 209--223.

\bibitem[{Seo \& Wolkowicz(2018)}]{seo2018sensitivity}
Seo, G. \& Wolkowicz, G.S. (2018).
\newblock Sensitivity of the dynamics of the general rosenzweig--macarthur
  model to the mathematical form of the functional response: a bifurcation
  theory approach.
\newblock \emph{Journal of Mathematical Biology}, 76, 1873--1906.

\bibitem[{Wiesenfeld(1985)}]{Wiesenfeld1985Noisy}
Wiesenfeld, K. (1985).
\newblock Noisy precursors of nonlinear instabilities.
\newblock \emph{Journal of Statistical Physics}, 38, 1071--1097.

\bibitem[{Wood \& Thomas(1999)}]{wood1999super}
Wood, S.N. \& Thomas, M.B. (1999).
\newblock Super--sensitivity to structure in biological models.
\newblock \emph{Proceedings of the Royal Society of London. Series B:
  Biological Sciences}, 266, 565--570.

\bibitem[{Wyse \emph{et~al.}(2022)Wyse, Martignoni, Mata, Foxall \&
  Tyson}]{wyse2022structural}
Wyse, S.K., Martignoni, M.M., Mata, M.A., Foxall, E. \& Tyson, R.C. (2022).
\newblock Structural sensitivity in the functional responses of predator--prey
  models.
\newblock \emph{Ecological Complexity}, 51, 101014.

\end{thebibliography}
\bibliographystyle{ecol_let}

\newpage

\renewcommand{\thesection}{Appendix~\Alph{section}}
\renewcommand{\theequation}{\Alph{section}\arabic{equation}}
\renewcommand{\thefigure}{\Alph{section}\arabic{figure}}

\setcounter{equation}{0}
\setcounter{section}{0}
\setcounter{figure}{0}

\section{Additional bifurcation diagrams}\label{app:add_bif_diags}

\begin{figure}[H]
    \centering
    \includegraphics[width=\textwidth]{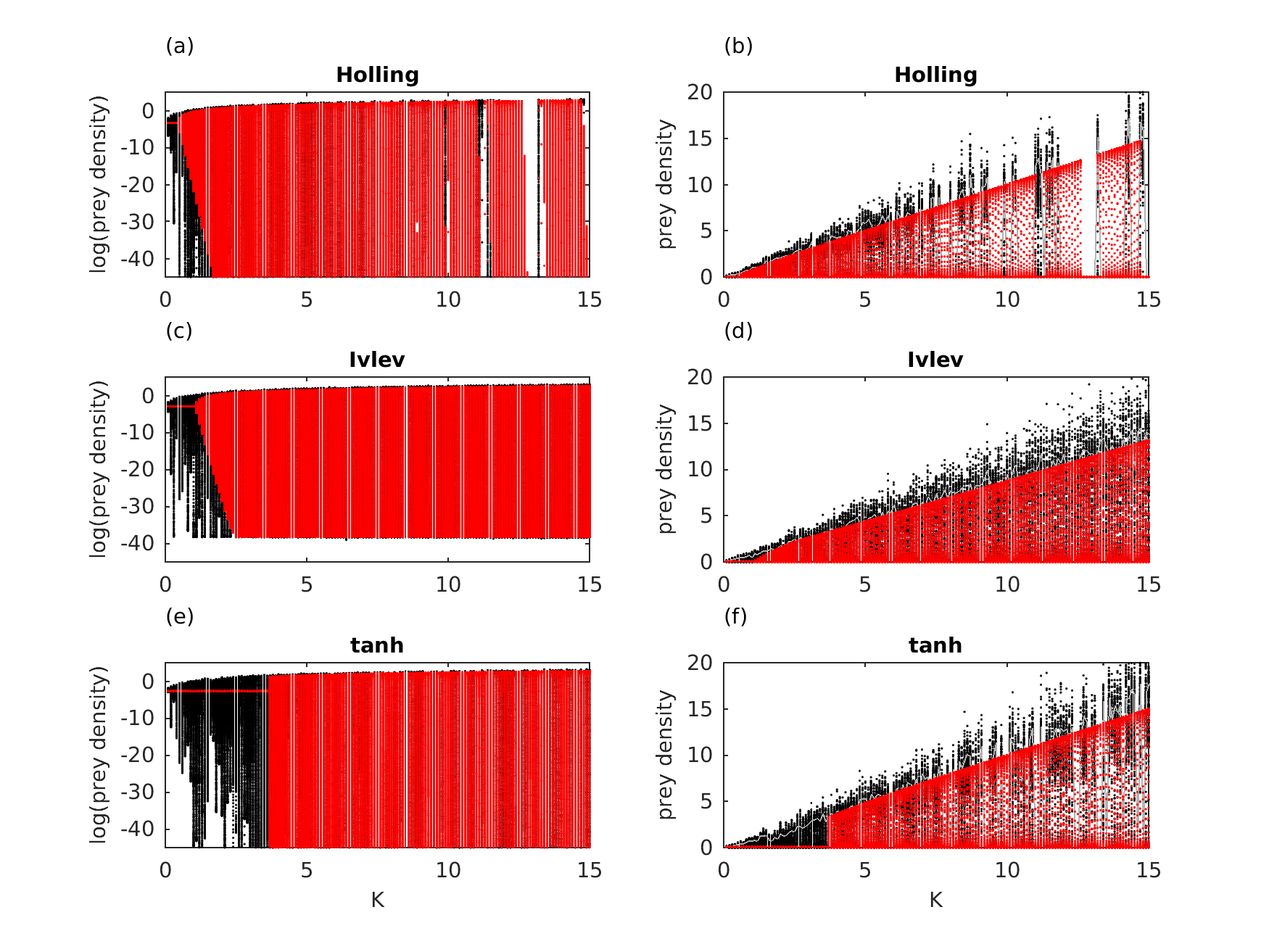}
    \caption{Bifurcation diagrams for the ODEs (red dots) and SDEs (black dots), up to large $K$, $\sigma_E=0.25$. Top (a-b), Holling; middle (b-c), Ivlev; bottom (e-f), tanh functional responses. Some extinction occurred in the Holling simulations for large K, hence the small gap; this might be remedied with finer-scale integration.}
    \label{fig:RMA_Kloop_largeK}
\end{figure}

We first zoom out to a large parameter space where $K$ is allowed to go up to 15 (Fig.~\ref{fig:RMA_Kloop_largeK}) as in \cite{fussmann2005community}, with the same variance of the noise as in the main text. The results are similar to those of the main text, that is, no sensitivity to model structure in the stochastic model. 

\begin{figure}[H]
    \centering
    \includegraphics[width=\textwidth]{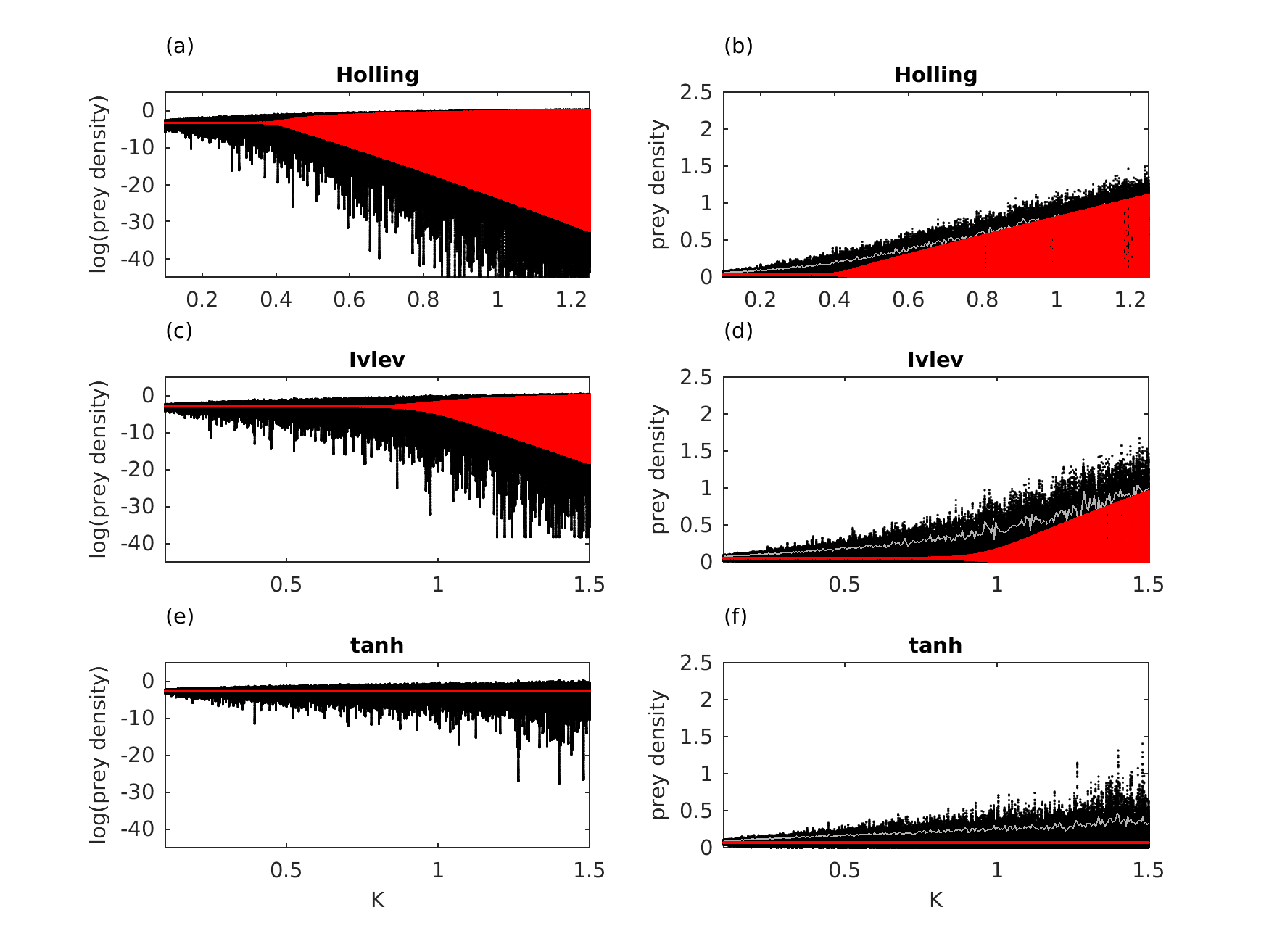}
    \caption{Bifurcation diagrams for the ODEs (red dots) and SDEs (black dots), $\sigma_E=0.1$. Top (a-b), Holling; middle (b-c), Ivlev; bottom (e-f), tanh functional responses. Other parameters otherwise identical to those of the main text.}
    \label{fig:RMA_Kloop_sigma01}
\end{figure}

Now we go back to small values of $K$ as in Fig.~\ref{fig:RMA_Kloop} of the main text, but we reduce the amount of stochasticity from $\sigma_E=0.25$ to $\sigma_E=0.1$ (Fig.~\ref{fig:RMA_Kloop_sigma01}). We see that the main result does hold---there is a smooth transition to large amplitude cycles for all three functional responses---although cycle amplitudes now differ a little between the three functional responses, especially with tanh. Still structural sensitivity has massively decreased with the introduction of stochasticity. 

\clearpage

\section{Power spectrum}\label{app:power_spectrum} 

\setcounter{equation}{0}
\setcounter{figure}{0}

\begin{figure}[H]
    \centering
    \includegraphics[width=0.95\textwidth]{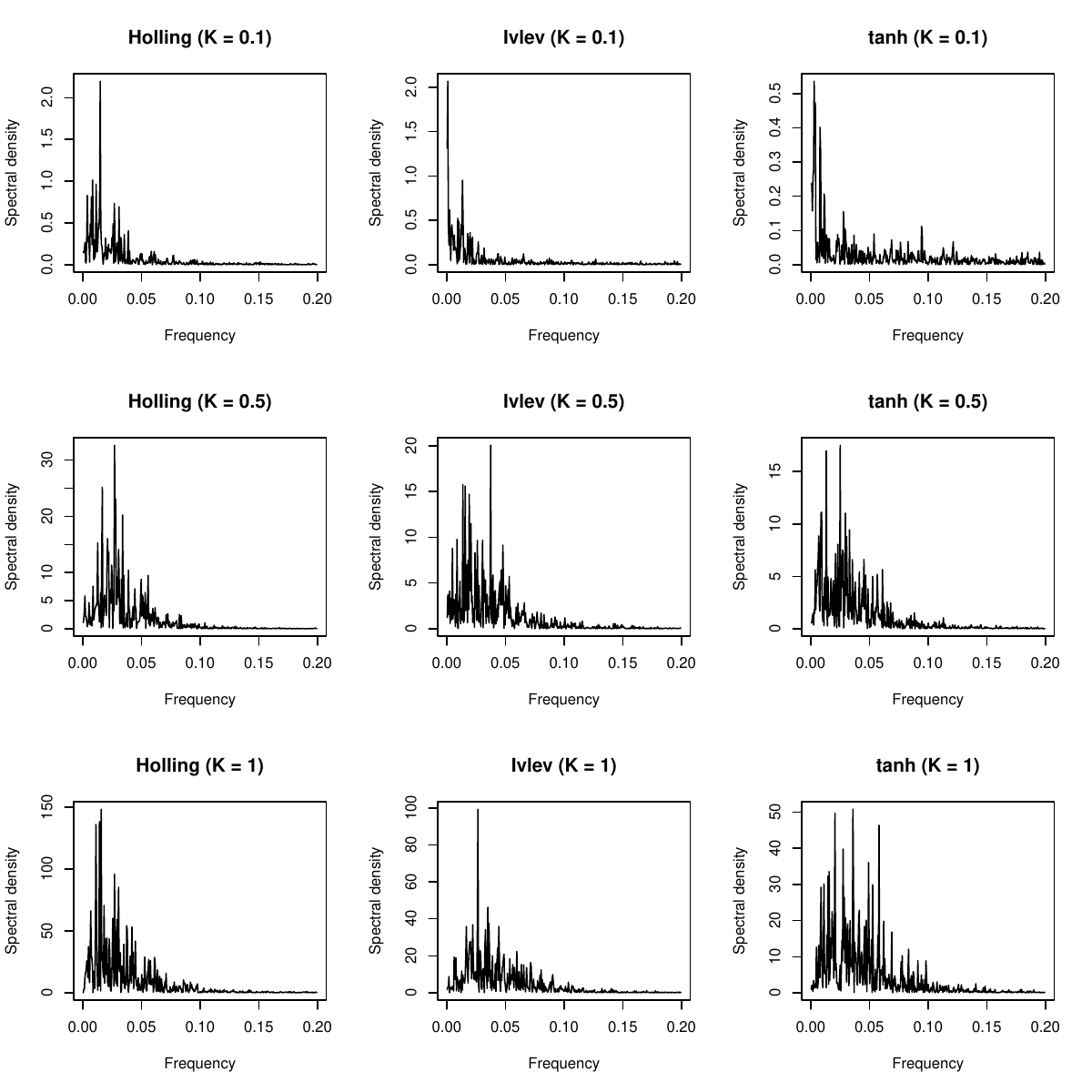}
    \caption{Power spectrum of $N$. Raw periodogram computed using the FFT with the function \texttt{spectrum} in \texttt{R}.} 
    \label{fig:spectrum}
\end{figure}

\begin{figure}[H]
    \centering
    \includegraphics[width=0.95\textwidth]{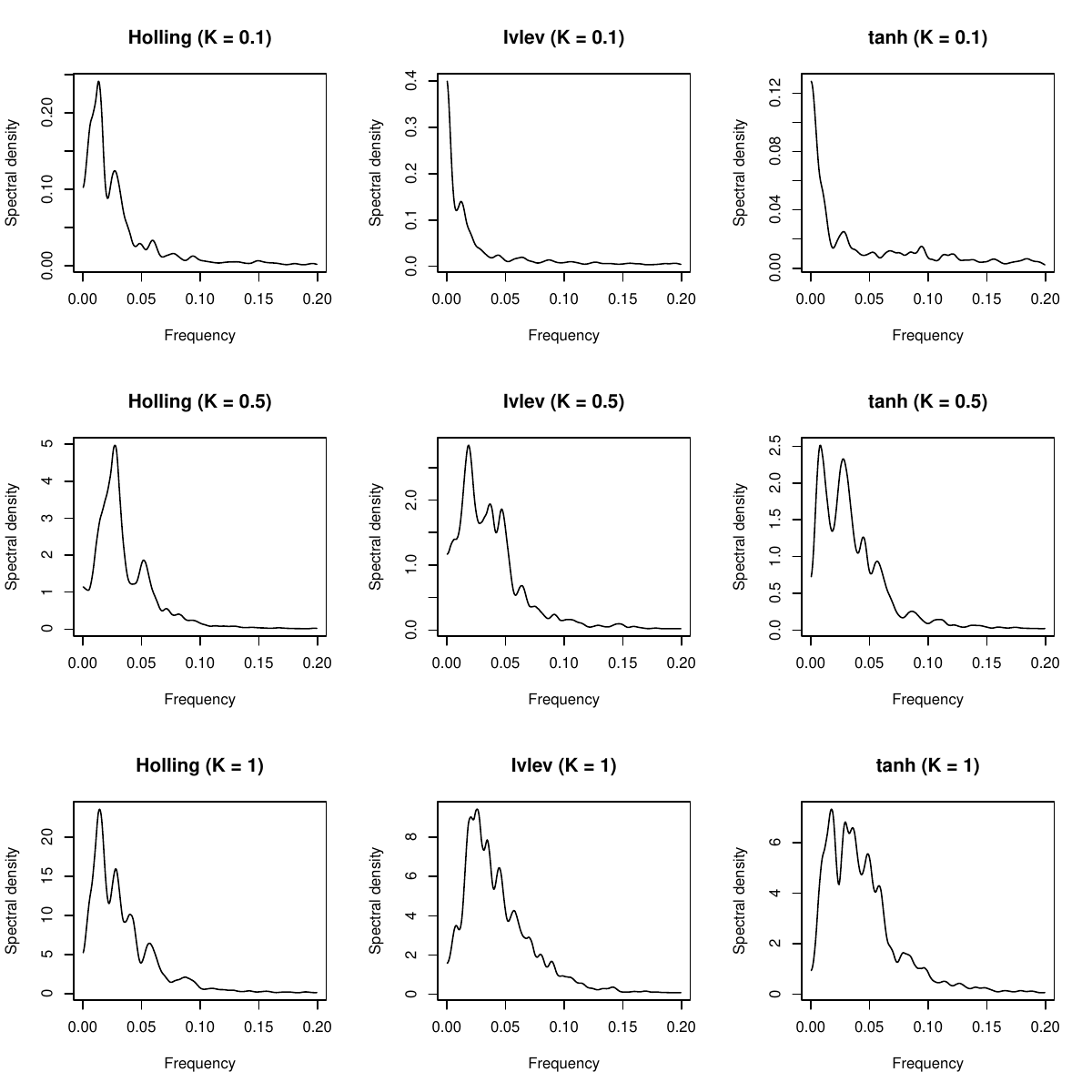}
    \caption{Power spectrum of $N$. Smoothed using a modified Daniell(11,11,11) kernel.}
    \label{fig:smoothed_spectrum}
\end{figure}

We present here the power spectrum for the time series of Fig.~\ref{fig:RMA_Traj_K0.5}, in a raw periodogram (Fig.~\ref{fig:spectrum}) and smoothed (Fig.~\ref{fig:smoothed_spectrum}) format. We see that the spectrum does not appear to be sensitive to details of the representation chosen. Although some differences appear in the spectrum for different values of $K$, no large differences between the three functional responses are noticeable (small differences at low frequencies should not be interpreted due to difficulties in smoothing the spectrum close to zero). Other techniques to reconstruct the Fourier spectrum (e.g. Welch, AR($p$)) tend to produce redder spectra and in some cases the dominant frequencies disappear, but (1) visual inspection of spacings between local maxima of the time series match the dominant frequencies presented here, which are below 1/20,  the frequency of quasi-cycles suggested by the linearized model, and (2) computation of the autocorrelation (logically) shows the same frequencies. A lower noise variance gets the dominant frequency closer to 1/20. 

\clearpage

\section{Large noise on the functional response}\label{app:large_noise_FR}

\setcounter{equation}{0}
\setcounter{figure}{0}

Here we make a stochastic functional response by introducing log-normal, autocorrelated noise on the half-saturation constant, a trick previously used by \cite{barraquand2021fitting} in a discrete-time context. The model now writes

\begin{align}
    dX = & \left( g(X) - f(X,Z) Y \right) dt + \sigma_E X dW_x\label{eq:SDE_RMA_noisyFR_prey}\\
    dY = & \left(\epsilon f(X,Z) Y -m Y \right) dt + \sigma_E Y dW_y
    \label{eq:SDE_noisyFR_RMA_noisyFR_pred}\\
    dZ = & - \gamma Z dt + \sigma_{\text{FR}} dW_z
    \label{eq:SDE_noisyFR_RMA_FR}
\end{align}

where the Holling type functional response is now 
\begin{equation}
    f_H(x,z) = \frac{C x}{D e^z + x}
\end{equation}
and the Ivlev-type 
\begin{equation}
    f_I(x,z)  = C \left( 1 - \exp \left(- \frac{\ln(2)}{D e^z} x \right) \right).
\end{equation}

\begin{figure}[H]
    \centering
    \includegraphics[width=\textwidth]{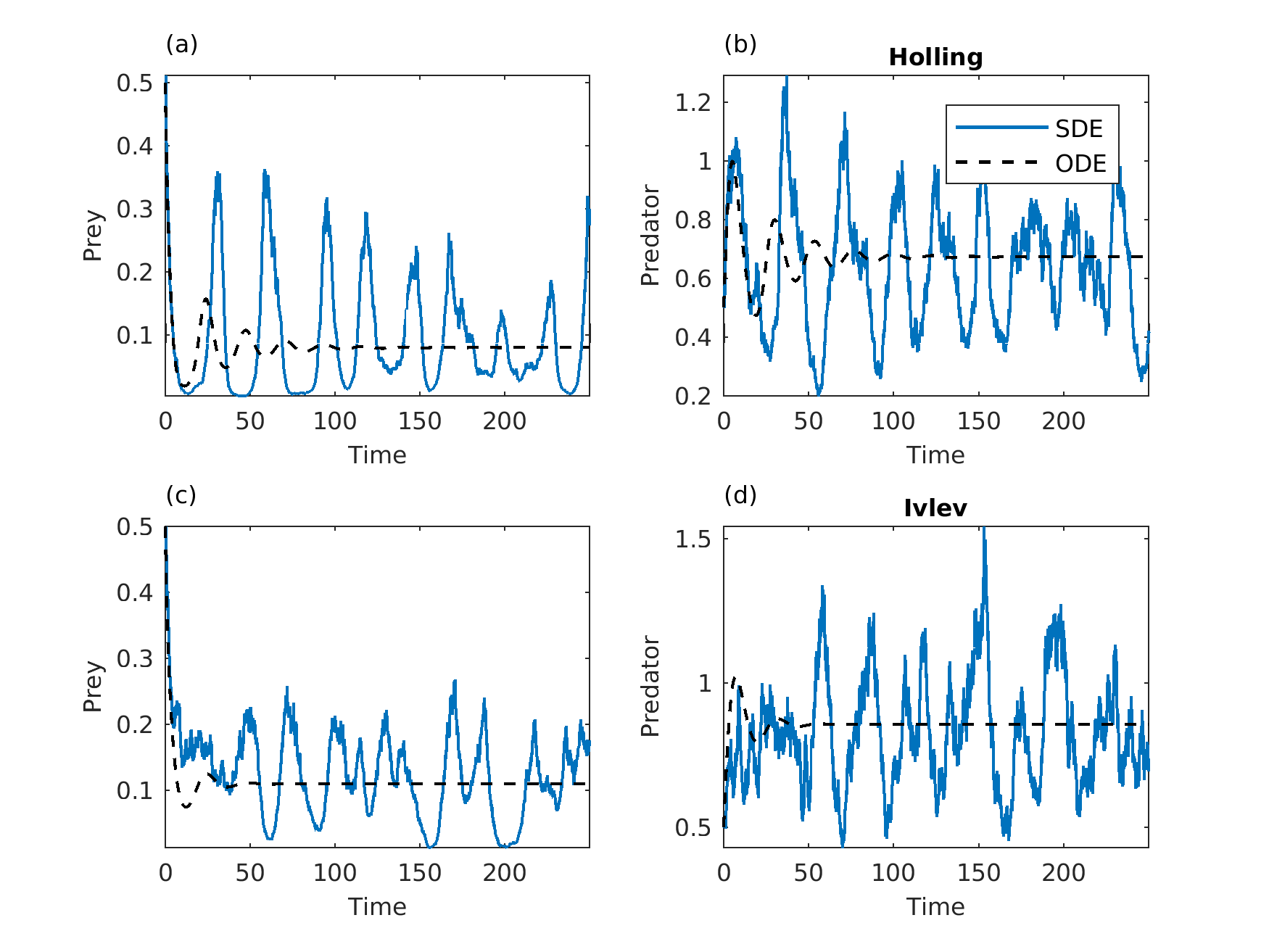}
    \caption{Trajectories of the ODEs (dashed black lines) and SDEs (plain blue lines) for 2 stochastic versions of the functional responses. Top (a-b), Holling; bottom (b-d), Ivlev. $K=0.5$. }
    \label{fig:RMA_noisy_FR_Traj_K05}
\end{figure}

\begin{figure}[H]
    \centering
    \includegraphics[width=\textwidth]{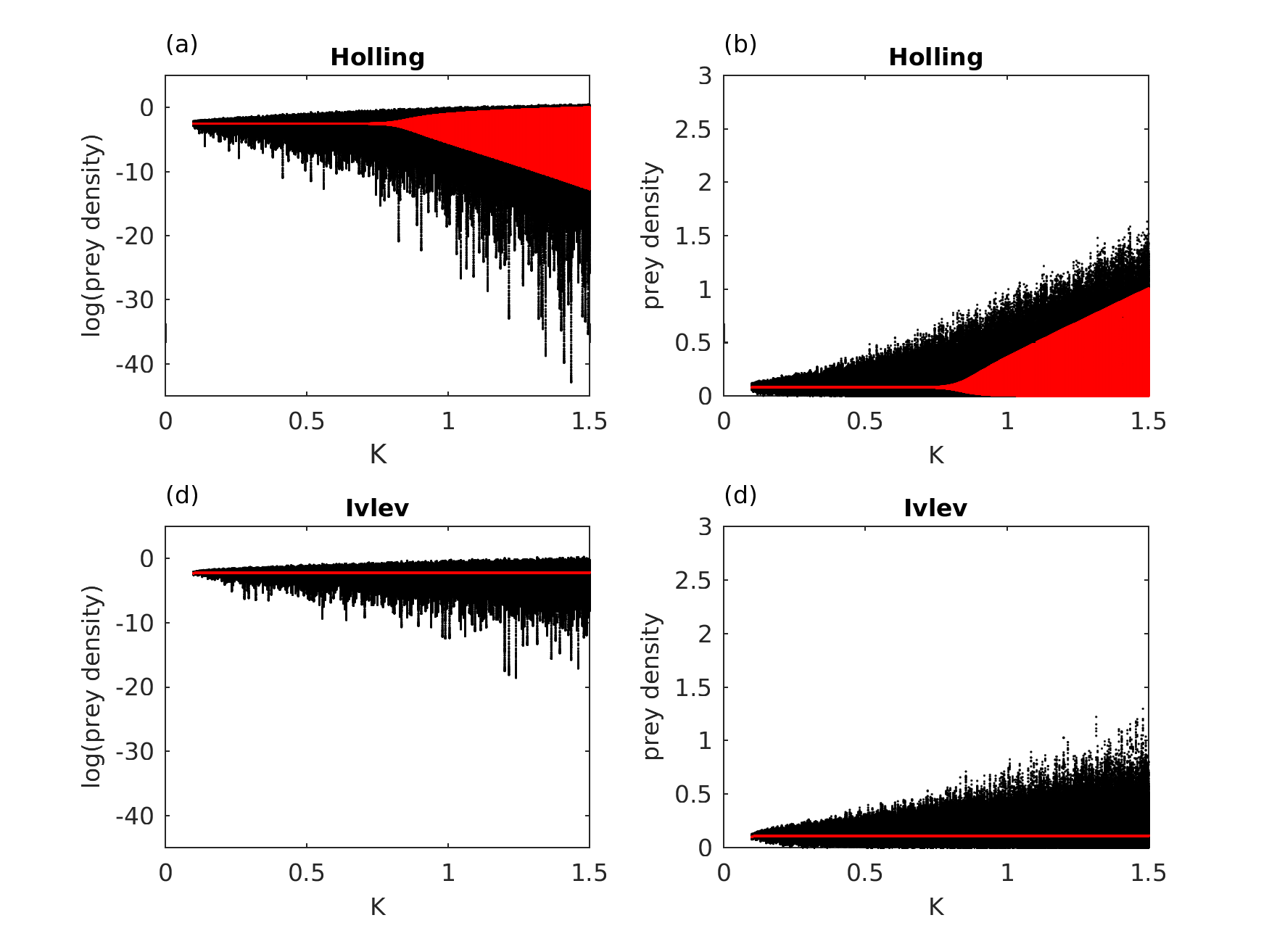}
    \caption{Bifurcation diagrams for the ODEs (red dots) and SDEs (black dots), $\sigma_E=0.1$ and $\sigma_{\text{FR}}=0.25$. Top (a-b), Holling; bottom (b-d), Ivlev functional response.}
    \label{fig:RMA_noisyFR_Kloop}
\end{figure}

The noisy functional responses generate time series shown in Fig.~\ref{fig:RMA_noisy_FR_Traj_K05}. We used $\sigma_E = 0.1$ and $\sigma_{\text{FR}}= 0.25$ to model moderate noise on the growth rates but large multiplicative noise on the functional response's half-saturation constant. We used $C = 1$ and $D = \frac{0.5}{\ln(2)}$ for both functional responses.
We only compare Holling and Ivlev functional responses for which the half-saturation is clearly defined as a parameter. Because it can be logical to assume that the functional response's half-saturation constant will not be fully white noise (it will vary with autocorrelated factors like other species' dynamics), we used an Ornstein-Uhlenbeck process, with parameter $\gamma = 0.1$. Noise on the functional response seem to make the dynamical system shift between alternate dynamical states, with small vs large oscillations, which might warrant further exploration (see also \cite{adamson2014defining} on this point). 
Although there is a little more amplitude in the Holling-type model, structural sensitivity has again massively decreased (see Fig. \ref{fig:RMA_noisyFR_Kloop}). 

\end{document}